\documentstyle[aps,pre,epsf,multicol]{revtex}
\newcommand{\be}{\begin{equation}}
\newcommand{\ee}{\end{equation}}
\newcommand{\ba}{\begin{array}}
\newcommand{\ea}{\end{array}}
\newcommand{\bea}{\begin{eqnarray}}

\newcommand{\eea}{\end{eqnarray}}
\newcommand{\del}{\nabla}
\newcommand{\ra}{\rightarrow}

\begin{document}
\draft
\title{General self-flattening surfaces}
\author{Hyunggyu Park}      
\address{School of Physics, Korea Institute for Advanced Study,
                 Seoul 130-722,  Korea}
%\date{\today}
\maketitle

\begin{abstract}
Recently Jeong and Kim [Phys.~Rev.~E {\bf 66}, 051605 (2002)] investigated 
the scaling properties of equilibrium self-flattening surfaces subject to a restricted curvature
constraint. In one dimension (1D), they found numerically that the stationary roughness exponent 
$\alpha\approx 0.561$ and the window exponent $\delta\approx 0.423$.
We present an analytic argument for general self-flattening surfaces in $D$ dimensions, leading to
$\alpha=D\alpha_0 /(D+\alpha_0)$ and $\delta=D/(D+\alpha_0)$ where
$\alpha_0$ is the roughness exponent for equilibrium surfaces without the
self-flattening mechanism. In case of surfaces subject to a restricted curvature
constraint, it is known exactly that $\alpha_0=3/2$ in 1D, which leads to
$\alpha=3/5$ and $\delta=2/5$. Small discrepancies between our analytic values and
their numerical values may be attributed to finite size effects. 

\end{abstract}


\begin{multicols}{2}

Fluctuation properties of equilibrium surfaces have been studied extensively
for the last few decades \cite{dyn}. Surface roughness is well documented and
classified into a few universality classes. The Edwards-Wilkinson (EW) class is
generic and robust for equilibrium surfaces with local surface tension \cite{EW}.
The EW surfaces can be described by the continuum Langevin-type equation 
\be
\frac{\partial h({\vec r}, t)}{\partial t} = - \nu \del^2 h({\vec r}, t) + \eta ({\vec r}, t),
\label{EWeq}
\ee
where $h({\vec r}, t)$ is the height at site ${\vec r}$ and time $t$,
$\eta({\vec r}, t)$ is an uncorrelated Gaussian noise, and $\nu$ 
represents the strength of local surface tension.

The surface fluctuation width $W(L,t)$, defined as the standard deviation of the 
surface height $h({\vec r}, t)$ starting from a flat surface of lateral size $L$,
satisfies the dynamic scaling relation
\begin{equation} 
W(L,t)=L^\alpha f\left({t / L^{z}}\right), \label{scaling} 
\end{equation}
where the scaling function $f(x)\ra {\rm const.}$ for $x\gg 1$ and 
$f(x) \sim x^\beta$ $(\beta =\alpha/z)$ for $x\ll 1$\cite{dyn,FAV}. 

The EW universality class is characterized by 
the dynamic exponent $z=2$ and the roughness exponent
$\alpha=(2-D)/2$ ($D$: substrate dimension).
In the absence of local surface tension $(\nu=0)$, higher-order local suppression terms 
like $\del^{2m} h$ $(m=2,3,\ldots)$ become relevant to determine the scaling 
properties of surface roughness. In this case, the scaling exponents become
$z=2m$ and $\alpha=(2m-D)/2$.

Recently Kim, Yoon, and Park \cite{KYP} introduced a global-type suppression
(self-flattening) mechanism which reduces
growing (eroding) probability only at the globally highest (lowest) point on the surface.
They found that this global-type suppression changes the scaling properties of
the EW surfaces: $z\approx 3/2$ and $\alpha=1/3$ in 1D and $z\approx 5/2$ 
and $\alpha=0(\log)$ in 2D. The 1D roughness exponent $\alpha=1/3$ characterizing
the stationary surface fluctuations could be derived exactly by mapping the
surface evolution model onto the static self-attracting random walk model \cite{SARW,DV}. 

More recently Jeong and Kim (JK) \cite{JK} investigated the effect of the self-flattening mechanism 
on the so-called restricted curvature (RC) model \cite{KDS}. The RC model is known to have no local
surface tension term $(\nu=0)$ and its dominant suppression term is of the fourth order 
$(\del^4 h)$. Accordingly, the ordinary RC model have the scaling exponents $z=4$ and
$\alpha = (4-D)/2$. Using Monte Carlo simulations, JK found for the 1D self-flattening
RC model that 
\begin{equation}
z=1.69(5),  \ \alpha=0.561(5), \ \mbox{and} \ \ \beta=0.332(5).
\label{RC-exp1}
\end{equation}
Again, the self-flattening mechanism changes the scaling properties of the RC surfaces.

JK also studied the height-height correlation function $G(r)$, defined as the average of
the square of height differences at two sites separated by a distance $r$. They found
an extra length scale $\xi$ (window size)  where the correlation function starts to saturate. 
The correlation function in the steady state satisfies the scaling relation 
\be
G(r)=L^{2\alpha} g\left({r/\xi}\right)  \ \  \mbox{with} \ \ \xi\sim L^\delta, \label{corr-func} 
\ee
where the scaling function $g(x)\ra {\rm const.}$ for $x\gg 1$ and 
$f(x) \sim x^{2\alpha^\prime-\kappa}$ $(\alpha^\prime =\alpha/\delta)$ for $x\ll 1$. 
This type of a crossover scaling, due to the existence of a smaller length scale 
$(\delta<1)$ compared to the system size, has been previously identified in the so-called 
even-visiting random walk (EVRW) models (see Eq.~(29) in reference \cite{NKPD}). 
In fact, the EVRW model is intimately related to the EW-type surface model with the self-flattening 
mechanism \cite{KYP}.
JK's numerical estimates for the exponents are 
\be
\delta\approx 0.423, \ \ \kappa\approx 0.868, \ \ \mbox{and} \ \ \alpha^\prime=\alpha/\delta\approx 1.33.
\label{RC-exp2}
\ee

In this Comment, we present an analytic argument that predicts the values of
the scaling exponents, $\alpha$ and $\delta$, 
associated with the stationary properties of surface fluctuations.
We consider a general equilibrium surface growth model, of which the stationary
roughness exponent is known exactly as $\alpha_0$. The partition function for
equilibrium self-flattening surfaces of lateral size $L$  can be written as \cite{KYP}
\begin{equation}
Z_L(K)=\sum_{\cal C} e^{-KH({\cal C})} ,
\label{partition}
\end{equation}
where the summation is over all possible height configurations ${\cal C}$
subject to a given constraint, $K$ is a temperaturelike parameter, 
and $H({\cal C})$ is the height excursion width (the globally maximum height minus 
the globally minimum height) for a given configuration ${\cal C}$.
Global suppression for self-flattening dynamics is simply Metropolis
type evolution algorithm with this partition function to reach the equilibrium \cite{KYP}. 
For the EW surfaces, one can take the surface height configurations subject to
the restricted solid-on-solid (RSOS) constraint, where the step heights are allowed
to take finite values. 
In the case of the RC surfaces, the local curvature $\del^2 h$ is restricted to be finite. 

One can decompose the configurational space into sectors with a constant excursion width $H$.
Then, the partition function can be rewritten as
\begin{equation}
Z_L(K)=\int_0^\infty dH\ \omega_L (H) \ e^{-KH} ,
\label{partition-2}
\end{equation}
where $\omega_L(H) dH$ is the number of configurations with the height excursion width 
between $H$ and $H+dH$. One can define $\Omega_L(H)$ 
as the number of configurations with the height excursion width less than $H$ as
\be
\Omega_L(H)=\int_0^H dH^\prime \ \omega(H^\prime),
\ee
and it is clear that $\Omega_L(\infty)=Z_L (K=0)$.

We estimate $\Omega_L (H)$ as follows. Consider a flat surface between 
two walls separated by a distance $H$ and parallel to the surface. The surface
starts to fluctuate, following its ordinary evolution dynamics. 
Whenever the surface hits
either of the two walls, the number of possible configurations inside the walls will be
reduced by a constant factor, compared to the no-wall case ($H=\infty$). 
This entropic reduction can be roughly translated as
\be
\Omega_L (H)\approx \Omega_L(\infty) \exp [-a N_c],
\ee
where $N_c$ is the typical number of contacts between the walls 
and the surface in the stationary state and $a$ is a positive constant of $O(1)$.
Typically, there will be $O(1)$ contacts over a block of lateral size $\ell$,
within which the stationary surface width $W_0 (\ell) \sim \ell^{\alpha_0}$
is of the same order of magnitude as  the wall distance $H$, i.e.,
\be
N_c\sim \left({L}/{\ell}\right)^D\sim {L^D}/{H^{D/\alpha_0}}.
\label{Nc}
\ee

With the above estimates, the partition function becomes for nonzero $K$
\begin{eqnarray}
Z_L(K)&=&Z_L(0) \int_0^\infty dH\    \left( \frac{\partial} {\partial H} 
e^{-aL^D/H^{D/\alpha_0}}\right) \ e^{-KH}, \nonumber \\
&=& Z_L(0) K \int_0^\infty dH\    e^{-KH-aL^D/H^{D/\alpha_0}}.
\end{eqnarray}
This integral can be evaluated by the saddle point method in the limit
of large $L$. The maximum contribution comes from 
\be
H^*\sim \left( \frac{L^D}{K}\right)^{\alpha_0/(D+\alpha_0)}.
\ee
The stationary  surface fluctuation width should be proportional to this typical
height excursion width, so we predict that
\be
W\sim  K^{-\alpha_0/(D+\alpha_0)}\ {L}^{\alpha},
\ee
with $\alpha=D\alpha_0/(D+\alpha_0)$.

The length scale $\ell$ arising naturally in our argument should be proportional
to the window size $\xi$ in the correlation function. Inside of this length scale,
the surface does not feel the global self-flattening suppression. 
As our length scale $\ell$ is explicitly related to the wall distance
in Eq.~(\ref{Nc}), we can also predict that
\be
\ell^*\sim \xi\sim L^\delta,
\ee
with $\delta=D/(D+\alpha_0)$.

In summary, for the general self-flattening equilibrium surfaces, we predict that
\be
\alpha=\frac{D\alpha_0}{D+\alpha_0}, \ \delta=\frac{D}{D+\alpha_0}, 
\ \ \mbox{and} \ \ \alpha^\prime=\frac{\alpha}{\delta}=\alpha_0,
\label{exp}
\ee
where $\alpha_0$ is the stationary roughness exponent for the equilibrium
surfaces without the self-flattening mechanism.

For the EW surfaces, $\alpha_0=1/2$ in 1D, which leads to $\alpha=1/3$ 
and $\delta=2/3$. These results agree with those by the other analytic 
(healing time) arguments \cite{NKPD,LD} and those by numerical simulations \cite{JKCK}. 
For the RC surfaces, $\alpha_0=3/2$ in 1D, which leads to $\alpha=3/5$ and $\delta=2/5$.
The short range behavior governing the value of $\kappa$ should be identical to the
ordinary RC model, so we also expect that $\kappa=1$.
These results have small discrepancies  from the JK's numerical results in
Eqs.~(\ref{RC-exp1}) and (\ref{RC-exp2}). We believe that this may be due to
rather small sizes used in their numerical simulations. The RC surfaces have $\alpha_0=(4-D)/2$ 
in $D$ dimensions, so $\alpha=D(4-D)/(4+D)$ and $\delta=2D/(4+D)$.
It may be interesting to check these predictions for the RC self-flattening surfaces in $D=2$ and 3.

At the upper critical dimensions ($D=2$ for the EW and $D=4$ for the RC surfaces),
the surface roughness becomes logarithmic ($\alpha_0 = 0$) and the self-flattening
mechanism induces only corrections to scaling in the surface fluctuations. The
dominant correction seems to be independent of system size $L$ \cite{KYP},
but its functional dependence on $K$ is not fully explored. 
In higher dimensions, the EW $(D>2)$and the RC $(D>4)$ surfaces are asymptotically flat
($\alpha_0<0$). The self-flattening mechanism induces power-law-type corrections to scaling
as given in Eq.~(\ref{exp}).

We could not present any analytic explanation for the dynamic exponents $z$ and $\beta$.
This may be due to the lack of a continuum-type equation to govern the self-flattening 
mechanism. It would be very interesting to find such an equation which should contain a global
coupling term in space. 

We thank J.~M.~Kim and H.-C.~Jeong for bringing us their work \cite{JK} in our attention. 
This work was supported by Grant No.~2000-2-11200-002-3 from the Basic Research Program of KOSEF.

\end{multicols}

\end{document}